\documentclass{svjour2}                    
\usepackage{graphicx}
\begin{document}

\title{Scattering a quantum particle on the potential step: Characteristic times for transmission and reflection}


\author{Nikolay L. Chuprikov 
}


\institute{N. L. Chuprikov \at
              Tomsk State Pedagogical University, 634041, Tomsk, Russia \\
              \email{chnl@tspu.edu.ru}           
}

\date{Received: date / Accepted: date}

\maketitle

\begin{abstract}
We present a new model of scattering a quantum particle on the potential step, which reconstructs the prehistory of the subensembles of
transmitted and reflected particles by their final states. Unlike the conventional one this model predicts the existence of a transitional spatial
region behind the potential step where the probability to find reflected particles is nonzero and the average velocity of transmitted particles
monotonously varies between asymptotic values. For both subprocesses we define the dwell and asymptotic group times. \keywords{dwell time \and
potential step \and group velocity } \PACS{03.65.-w\ 03.65.Xp \ 42.25.Bs}
\end{abstract}

\newcommand{\ppp}{\mbox{\hspace{5mm}}}
\newcommand{\ooo}{\mbox{\hspace{3mm}}}
\newcommand{\ooa}{\mbox{\hspace{1mm}}}
\newcommand{\ppd}{\mbox{\hspace{18mm}}}
\newcommand{\ppt}{\mbox{\hspace{34mm}}}
\newcommand{\ppo}{\mbox{\hspace{10mm}}}

\section{Introduction}

As far as we know, to date there are at least two approaches \cite{Leo,Dav} in the tunneling time literature (TTL) that deal directly with the
temporal aspects of scattering a non-relativistic quantum particle by the potential step (see also the recent paper \cite{Xue} where the approach
\cite{Dav} is generalized onto relativistic particles). Both approaches are treated in the TTL as realistic ones since they define the
transmission and reflection times with making use of clock models which imply a direct measurement of these scattering times.

Our purpose is to show that these models violate the causality principle and their results can hardly be called realistic. We begin with the most
striking result whose essence is the following (see \cite{Dav}). When the particle's energy is less by an infinitesimal magnitude than the height
of the potential step (situated for example at the point $x=a$) the average time of staying a particle in the spatial region $x>a$ located behind
the step diverges; that is, in this case the particle takes (on average) an infinite time to bounce back into the region $(-\infty,a]$. However,
when its energy is higher by an infinitesimal magnitude than the step height, the particle bounces off the step instantaneously.

Davies emphasizes the paradoxicality of this result. But he believes that it creates no interpretational problems since this discontinuity occurs
just at that point on the energy scale where the clock model \cite{Dav} is unreliable because of its limited resolution.

But such an explanation is not convincing since the reflection time defined in this model exhibits strange physical properties not only at this
point. This occurs at all energies exceeding the step height: according to this model, the reflection time is nonzero when the particle's energy
is less than the step height; whilst otherwise it equals strictly to zero. That is, this model says, in fact, that the probability to find
reflected particles in the spatial region behind the step is nonzero when this region is classically forbidden, but zero when this region is
classically accessible.

Note that this consequence of \cite{Dav} is in a full agreement with the conventional quantum-mechanical model (CQM) of scattering a particle on
the potential step. Under this model, when the particle energy is less by an infinitesimal magnitude than the step height, the probability to find
reflected particles, at a finite distance behind the step, is nonzero. However, if its energy exceeds the step height by an infinitesimal
magnitude, then this probability equals strictly to zero everywhere in this region (where only the transmitted wave exists).

That is, according to the CQM, in the first case, reflected particles are able to penetrate (due to their wave properties) into the classically
forbidden spatial region. While in the second case (when this region becomes classically accessible, and the difference in energy, in the
comparison with the first case, is infinitesimal) they cannot. Such energy dependence of the probability density, in the CQM, is evident to
contradict the very spirit of quantum mechanics. So that the strange behavior of the reflection time in the clock model \cite{Dav}, based fully on
the CQM, is a consequence of the latter.

Another argument against the CQM is that it, in principle, does not allow one to introduce the transmission and reflection times in accordance
with the causality principle. To show this, let us analyze in detail the presented in \cite{Dav} idea of introducing the transmission time:
\begin{quote}
"To achieve this, the particle is coupled (weakly) to a quantum clock. The coupling is chosen to be non-zero only when the particle's position
lies within a given spatial interval\ldots Initially the clock pointer is set to zero. After a long time, when the particle has traversed the
spatial region of interest with high probability, the position of the clock pointer is measured. The change in position yields the expectation
value for the time of flight of the particle between the two fixed points."
\end{quote}

Note that this clock model has been designed for timekeeping a {\it scattering} particle, but its key features have been illustrated in \cite{Dav}
by the example of a {\it free} particle. That is, per se, the dynamics of free and scattering particles are treated in \cite{Dav} as qualitatively
identical. But this is not; this model, being true in the case of a free particle, violates the causality principle in the case of a particle
scattering by the potential step or rectangular barrier.

To see the principal difference between the timekeeping of the quantum dynamics of free and scattering particles, one has to take into account the
following two points. First, the mentioned coupling between a quantum clock and a particle in the stationary state is realized in \cite{Dav} as a
coupling between the clock and the phase of the wave function that describes the particle's state. Second, when "the spatial region of interest"
is empty the incident wave, that traverses it, never splits into parts. Whilst, when this region contains the potential step or rectangular
barrier, the incident wave splits here into two waves -- transmitted and reflected.

The latter means that at the initial stage of scattering the clock pointer is coupled in this model to the phase of the {\it incident} wave (that
describes the whole ensemble of particles, without distinguishing its to-be-transmitted and to-be-reflected parts), while at the final stage it is
coupled to the phase of the {\it transmitted} wave (that describes only the transmitted part of the incident ensemble). That is, in this model,
the procedure of setting to zero of the clock pointer is based, in fact, on the implicit assumption that the average departure times of the whole
ensemble of particles and its to-be-transmitted part (causally connected to the transmitted subensemble), coincide with each other. But this is
not (see \cite{La2}). So that the clock model \cite{Dav} violates the causality principle. To correctly perform this procedure for clocks used for
measuring the transmission (reflection) time, one needs first to restore the whole prehistory of the subensemble of transmitted (reflected)
particles by the transmitted (reflected) wave.

One more crucial point, that should be taken into account in studying the temporal aspects of both subprocesses, is that the transmission and
reflection times, even being properly defined, cannot be {\it directly} measured because of interference between these subprocesses. Neither the
measurement of these characteristic times nor the unambiguous interpretation of measurement data can be properly realized without a
quantum-mechanical model of the scattering process, that would able to 'see' the dynamics of each subprocess at all stages of scattering. In
particular, the key point of all such measurements -- zero setting of the clock pointer -- cannot be correctly performed if the dynamics of each
subprocess at the initial stage of scattering is unknown.

In the papers~\cite{Ch2,Ch3,Ch4,Ch5} we suggest, by the example of scattering a particle by symmetric potential barriers, the technique of
restoring the individual prehistory of the subensembles of transmitted and reflected particles by their final states. The main idea of this
approach is to present the total wave function that describes this process as a superposition of two "subprocess wave functions" (SWFs) that
describe the dynamics of each subprocess at all stages of scattering. According to this approach, any characteristic time and expectation value of
any one-particle physical observable can be calculated {\it only} for subprocesses.

This idea contradicts the conventional practice of a quantum-mechanical description of scattering a particle by 1D potential barriers. And its
proponents claim that any decomposition into SWFs of the total wave function -- the solution of the scattering problem, continuous everywhere
together with its first spatial derivative -- is inadmissible. Only this wave function they say should be used for calculating any physical
quantities; no part of the total wave function can be used for that.

However, as was shown above, characteristic times for the potential step cannot be correctly calculated on the basis of the total wave function.
Moreover, even the transmission and reflection coefficients cannot be calculated on its basis: the incident wave that is used for this purpose is
just a part of the total wave function. Besides, for the final stage of a time-dependent scattering, it is clearly pointless to calculate the
expectation values of the particle's position and momentum on the basis of the total wave function, because it represents at this stage the
superposition of the transmitted and reflected wave packets occupying macroscopically distinct spatial regions.

So that the validity of this practice is an illusion, in fact, and the idea to apply the decomposition technique \cite{Ch2} to the potential step
is quite motivated.

\section{Backgrounds}\label{Back}

We begin with the CQM. Let a particle with a given momentum $\hbar k$ impinges from the left on the potential step of height $V_0$, situated at
the point $x=a$ ($a>0$). The total wave function $\Psi_{tot}(x,k)$ to describe this scattering process when $E>V_0$, where $E=\hbar^2 k^2/2m$, is
\begin{eqnarray} \label{eq.1}
\Psi_{tot}(x,k)=\left\{\begin{array}{rl}
e^{ikx}+B e^{-ikx} \ppp x<a \\
A e^{i\kappa x} \ppp x>a
\end{array} \right. \\
A=\frac{2k}{k+\kappa}e^{i(k-\kappa)a},\ppp B=\frac{k-\kappa}{k+\kappa}e^{2i k a};\nonumber\\
\kappa=k\sqrt{1-\beta (\kappa_0/k)^2},\ooa \kappa_0=\sqrt{2m|V_0|}/\hbar,\ooa \beta=sgn(V_0);\nonumber
\end{eqnarray}
the corresponding transmission and reflection coefficients are
\begin{eqnarray*}
T=\frac{4k\kappa}{(k+\kappa)^2},\ppp R=\left(\frac{k-\kappa}{k+\kappa}\right)^2.
\end{eqnarray*}

The wave function $\Psi_{tot}(x,t)$ to describe a time-dependent scattering process is
\begin{eqnarray} \label{100}
\Psi_{tot}(x,t)=\frac{1}{\sqrt{2\pi}} \int_{-\infty}^\infty \mathcal{A}(k) \Psi_{tot}(x,k) e^{-iE(k)t/\hbar}dk;
\end{eqnarray}
$\mathcal{A}(k)$ is the Gaussian function $\mathcal{A}(k)=(2l_0^2/\pi)^{1/4}\exp[-l_0^2(k-k_0)^2]$ for example. In this setting of the problem, at
the initial time $t=0$
\begin{eqnarray} \label{101}
\bar{x}_{tot}(0)=0,\ppp \bar{p}_{tot}(0)=\hbar\bar{k},\ppp\overline{x^2}_{tot}(0)=l_0^2;
\end{eqnarray}
hereinafter, for the average value of any physical observable $F$ and localized state $\Psi^A_B$ we will use two equivalent notations: \[
\bar{F}^A_B(t)\equiv <F>^A_B(t) =\frac{<\Psi^A_B|\hat{F}|\Psi^A_B>}{<\Psi^A_B|\Psi^A_B>};\] if $\bar{F}^A_B(t)$ is constant its argument will be
omitted. We assume that the parameters $l_0$ and $\bar{k}$ obey the conditions that are necessary for a one-dimensional {\it completed} scattering
(OCS): the rate of scattering the transmitted and reflected wave packets exceeds the rate of widening each packet, so that the transmitted and
reflected wave packets non-overlap each other at the final stage of scattering, i.e., when $t\to\infty$. We also assume that the origin of
coordinates, which is the starting point of the "center of mass" (CM) $\bar{x}_{tot}(t)$ of the wave packet $\Psi_{tot}(x,t)$, lies far enough
from the potential step: $a\gg l_0$.

At the initial stage of the OCS, i.e., long before the scattering event,
\begin{eqnarray}\label{103}
\Psi_{tot}(x,t)\simeq \Psi_{inc}(x,t)=\frac{1}{\sqrt{2\pi}}\int_{-\infty}^{\infty} \mathcal{A}(k)\ooa e^{i[kx-iE(k)t/\hbar]}dk.
\end{eqnarray}
At the final stage of the OCS, i.e., long after the scattering event,
\begin{eqnarray}\label{104}
\Psi_{tot}(x,t)= \Psi_{tr}(x,t)+\Psi_{ref}(x,t);
\end{eqnarray}
\begin{eqnarray*}
\Psi_{tr}(x,t)=\frac{1}{\sqrt{2\pi}}\int_{-\infty}^{\infty}\mathcal{A}(k) A(k)\ooa e^{i\left[\kappa
x-E(k)t/\hbar\right]}dk,\\
\Psi_{ref}(x,t)=\frac{1}{\sqrt{2\pi}}\int_{-\infty}^{\infty}\mathcal{A}(k)B(k)\ooa e^{-i\left[kx+E(k)t/\hbar\right]}dk.
\end{eqnarray*}

Our next step is to present the scattering process described by $\Psi_{tot}(x,t)$ as a complex process consisting of two alternative subprocesses,
transmission and reflection, described at all stages of scattering by the SWFs $\psi_{tr}(x,t)$ and $\psi_{ref}(x,t)$, respectively.

\section{The total wave function as a superposition of two SWFs} \label{ref_tr}
\subsection{Physical requirements imposed on the SWFs}

According to \cite{Ch2}, the uniqueness of partitioning this time-dependent scattering process into subprocesses, for any function
$\mathcal{A}(k)$, is provided by the uniqueness of presenting the total stationary state $\Psi_{tot}(x,k)$ in the form of a superposition of the
SWFs $\psi_{tr}(x,k)$ and $\psi_{ref}(x,k)$. The latter, in its turn, is provided by physically motivated requirements imposed on the SWFs.

The first requirement follows from the fact that at the final stage of scattering the total wave function $\Psi_{tot}(x,t)$ represents the
superposition of two nonoverlapping wave packets, transmitted and reflected (see (\ref{104})). Thus, to provide the fulfilment of this property
for any function $\mathcal{A}(k)$, the stationary SWFs must be such that, for any values of $x$ and $k$,
\begin{itemize}
\item[(a)] $\psi_{tr}(x,k)+\psi_{ref}(x,k)=\Psi_{tot}(x,k)$.
\end{itemize}
In this case, the time-dependent SWFs
\begin{eqnarray} \label{200}
\psi_{tr,ref}(x,t)=\frac{1}{\sqrt{2\pi}} \int_{-\infty}^\infty \mathcal{A}(k) \psi_{tr,ref}(x,k) e^{-iE(k)t/\hbar}dk
\end{eqnarray}
are evident to obey the equality
\begin{eqnarray}\label{201}
\Psi_{tot}(x,t)=\psi_{tr}(x,t)+\psi_{ref}(x,t).
\end{eqnarray}

Next requirement is dictated by the very nature of the SWFs:
\begin{itemize}
\item[(b)] each SWF must have only one incoming wave and only one outgoing wave; the outgoing wave of $\psi_{tr}(x,k)$ is $A
e^{i\kappa x}$ and that of $\psi_{ref}(x,k)$ is $B e^{-ikx}$ (see (\ref{eq.1})).
\end{itemize}
As is known, for $E>V_0$ there is no solution to the stationary Schr\"{o}dinger equation, which would be everywhere continuous together with its
first spatial derivative and simultaneously possess one incoming wave and one outgoing wave. Thus, we must so weaken the continuity conditions of
the SWFs that they would ensure, on the one hand, the existence of a nontrivial resolution of the decomposition problem and, on the other hand, a
causal relationship between the incoming and outgoing waves in each SWF.

Namely, we believe that an incoming wave and outgoing wave of each stationary SWF represent two different solutions to the Schr\"{o}dinger
equation (for the same potential step), that are joined together at some spatial point $x_c(k)$ according to the following requirement:
\begin{itemize}
\item[(c)] at the point $x_c(k)$ where the incoming and outgoing waves of each stationary SWF are joined to each other,
the SWF must be continuous together with the corresponding probability current density.
\end{itemize}

According to the (b) and (c) requirements, the SWF $\psi_{ref}(x,k)$ must be zero in the region $x>x_c(k)$ because its incoming and outgoing waves
move to the left of the point $x_c(k)$. Thus, $\psi_{ref}(x,k)$ is a currentless wave function and the point $x_c(k)$, that plays in this model
the role of an extreme right turning point for reflected particles, is evident to coincide with some zero of this function. The right zero obeys
the following requirement:
\begin{itemize}
\item[(d)] for any given value of $k$ the point $x_c(k)$ must coincide with such a zero of the currentless wave function $\psi_{ref}(x,k)$,
at which the absolute value of $dx_c(k)/dk$ is less than at other zeros of this function.
\end{itemize}
The crucial difference between the $k$ dependence of the sought-for zero and others is most noticeable in the limiting cases $k\to\kappa_0+0$ when
$V_0>0$, and $k\to 0$ when $V_0<0$. As will be seen from the following, for the step potential the point $x_c(k)$ always lies behind the step,
i.e., in the region $x>a$. Thus, according to our approach, the interval $[a,x_c(k)]$ is accessible for reflected particles; hereinafter, it will
be referred to as 'the transitional region'. Note that, according to the CQM, reflected particles can enter the region $x>a$ only when $V_0>0$ and
$E<V_0$ (see Introduction and Section \ref{compref}).

Our next step is to realize the above program of decomposing the total wave function into SWFs.

\subsection{Wave functions for transmission and reflection}

Note that, due to the (b) and (c) requirements, the search of the SWFs $\psi_{tr}(x,k)$ and $\psi_{ref}(x,k)$ reduces in fact to finding the
amplitudes of their incoming waves as well as to finding the position of the point $x_c$. In this case the (a), (b) and (c) requirements uniquely
determine these amplitudes. Indeed, according to the (b) and (c) requirements, both SWFs in the region $x<a$ can be written in the form
\begin{eqnarray} \label{eq.2}
\psi_{tr}(x,k)=A_{tr}e^{ikx},\ppp \psi_{ref}(x,k)=A_{ref}e^{ikx}+B e^{-ikx};
\end{eqnarray}
and in the region $x>x_c(k)$ they are
\begin{eqnarray} \label{203}
\psi_{tr}(x,k)=A e^{i\kappa x},\ppp \psi_{ref}(x,k)=0.
\end{eqnarray}
As the probability current densities of the incoming and outgoing waves of each SWF are equal,
\begin{eqnarray} \label{204}
|A_{tr}|^2=\frac{\kappa}{k}\ooa |A|^2=T,\ppp |A_{ref}|^2=|B|^2=R.
\end{eqnarray}
Thus, the complex amplitudes $A_{tr}$ and $A_{ref}$ can be written in the form $A_{tr}=\sqrt{T}e^{i\mu}$ and $A_{ref}=\sqrt{R}e^{i\nu}$, where
$\mu$ and $\nu$ are unknown phases. Since $A_{tr}+A_{ref}=1$, what follows from the (a) requirement, these phases obey the equation
$\sqrt{T}e^{i\mu}+\sqrt{R}e^{i\nu}=1$. Solving this equation for any given value of $k$, we obtain two pairs of roots:
\begin{eqnarray} \label{206}
\mu_1=\lambda- \frac{\pi}{2},\ooo \nu_1=\lambda;\ppp\ppp \mu_2=-\lambda+\frac{\pi}{2},\ooo \nu_2=-\lambda;\nonumber\\
\lambda=\arctan{\sqrt{T/R}}=2\arctan{\sqrt{(\kappa/k)^\beta}}.
\end{eqnarray}

However, only the pair of amplitudes
\begin{eqnarray} \label{209}
A_{ref}=\sqrt{R}\exp(i \beta \lambda),\ppp  A_{tr}=\sqrt{T}\exp\left[i \beta \left(\lambda-\frac{\pi}{2}\right)\right].
\end{eqnarray}
gives $\psi_{ref}(x,k)$ whose zero in the region $x>a$, nearest to the step, obeys the (d) requirement and hence can serve as the extreme right
turning point $x_c$. For $x<a$ this function is determined by the expression (see also (\ref{eq.2}))
\begin{eqnarray} \label{208}
\psi_{ref}(x,k)=\sqrt{R}e^{i(ka+\beta\lambda/2)}\left[e^{i[k(x-a)+\beta\lambda/2]}+\beta e^{-i[k(x-a)+\beta\lambda/2]}\right].
\end{eqnarray}
In the transitional region $[a,x_c(k)]$ it can be written in the form
\begin{eqnarray} \label{210}
\psi_{ref}(x,k)=C \sin\left[\kappa(x-x_c)\right];\\
C=-2i^{\frac{\beta-1}{2}}\sqrt{\frac{k}{\kappa}R}\ooa\exp\left[i\left(ka+\beta\frac{\lambda}{2}\right)\right],\ooo
x_c=a+\frac{1}{\kappa}\arctan\sqrt{\frac{\kappa}{k}}\ooa . \nonumber
\end{eqnarray}

Note that in the limiting cases we have (see also fig.~\ref{fig.3})
\begin{itemize}
\item[$\bullet$] $V_0>0$ and $k\to\kappa_0+0$ ($\kappa\to 0$): $(x_c-a)\to (\kappa\kappa_0)^{-1/2}$ and $\kappa(x_c-a)\to 0$;
\item[$\bullet$] $V_0<0$ and $k\to 0$: $(x_c-a)\to \pi/(2\kappa_0)$ and $\kappa(x_c-a)\to \pi/2$;
\item[$\bullet$] $k\to\infty$: $(x_c-a)\to +0$ and $\kappa(x_c-a)\to\pi/4$ for both signs of $V_0$.
\end{itemize}
As is seen, when $V_0>0$ the length $x_c-a$ tends to infinity as $\kappa^{-1/2}$ in the limit $\kappa\to 0$. Note that $\kappa(x_c-a)\to 0$ in
this case! For all other zeros, the corresponding lengths increase in this limit as $\kappa^{-1}$.

Thus, according to our model, the probability to find reflected particles to the right of the step is nonzero for both signs of $V_0$. And this
takes place in the transitional region $[a,x_c(k)]$. The CQM does not imply the existence of such a region. This model says that for $V_0>0$ the
penetration depth of reflected particles is infinite for $E\to V_0-0$ but zero for $E\to V_0+0$ (see \cite{Dav}). As it will be shown below, in
our model this quantity is infinite in both these limits (see Exps. (\ref{eq.51}) and (\ref{eq.12}), as well as fig~\ref{fig.4}).

Note also that for the alternative pair of roots in (\ref{206}) the corresponding function $\psi_{ref}^{alt}(x,k)$ is nonzero at the point $x_c$;
but its first spatial derivative is zero here. For both signs of $V_0$, there is no zero of $\psi_{ref}^{alt}(x,k)$ which would lie between the
step $x=a$ and zero $x_c(k)$ of the function $\psi_{ref}(x,k)$.

\section{Characteristic times for transmission and reflection}

Since the dynamics of both subprocesses is now known in all spatial regions, we can proceed to studying their temporal aspects. Our aim is to
define for each subprocess the dwell and asymptotic group times. The former will be defined on the basis of the flow velocity concept for the
stationary scattering problem. The latter will be defined on the basis of the group velocity concept for narrow in $k$ space wave packets (the
value of $l_0$ is assumed to be large enough).

\subsection{Dwell times for transmission and reflection}

\begin{figure}
\begin{center}
\includegraphics[scale=0.2]{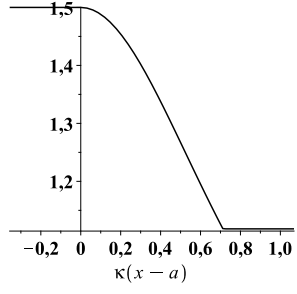}
\caption{The relative transmission velocity $v_{tr-flow}/v_0$ for $V_0>0$ and $k/ \kappa_0=1.5$; $v_0=\hbar \kappa_0/m$} \label{fig.1}
\end{center}
\end{figure}

\begin{figure}[t]
\begin{center}
\includegraphics[scale=0.2]{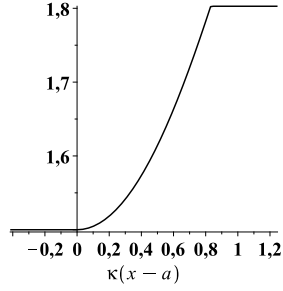}
\caption{The relative transmission velocity $v_{tr-flow}/v_0$ for $V_0<0$ and $k/ \kappa_0=1.5$} \label{fig.2}
\end{center}
\end{figure}

In the case of transmission the flow velocity $v_{tr-flow}(x,k)$ at any point $x$ is
\begin{eqnarray} \label{300}
v_{tr-flow}(x,k)=\frac{I_{tr}}{|\psi_{tr}(x,k)|^2};\ppp I_{tr}=I_{tot}=\frac{\hbar k}{m} T(k).
\end{eqnarray}
Figs.~\ref{fig.1} and \ref{fig.2} show that in the regions $x\leq a$ and $x\geq x_c$ the flow velocity of transmitted particles coincides with the
'free-particle' velocities $v_k=\hbar k/m$ and $v_\kappa=\hbar \kappa/m$, respectively. In the transitional region $[a,x_c]$ it varies
monotonously between these two asymptotic values. When $V_0>0$ ($V_0<0$), the average velocity of transmitted particles in this region is higher
(smaller) than $v_\kappa$. Thus, since a particle is free (in the above sense) outside the transitional region, it is sufficient to consider in
detail only this region.

The transmission dwell time for this region is defined as
\begin{eqnarray*}
\tau_{tr-dwell}=\frac{1}{I_{tr}}\int_a^{x_c}|\psi_{tr}(x,k)|^2dx.
\end{eqnarray*}
As regards reflected particles, the flow velocity concept is inapplicable for revealing their average velocity in the transitional region because
$I_{ref}=0$. Therefore the reflection dwell time for the transitional region is defined here in line with \cite{But}:
\begin{eqnarray*}
\tau_{ref-dwell}=\frac{1}{I_{ref}^{inc}}\int_a^{x_c}|\psi_{ref}(x,k)|^2dx;\ppp I_{ref}^{inc}=\frac{\hbar k}{m}R(k).
\end{eqnarray*}
Considering the equality $\psi_{tr}(x,k)=\Psi_{tot}(x,k)-\psi_{ref}(x,k)$ and Exps. (\ref{210}) for $\psi_{ref}(x,k)$, we obtain dwell times for
both subprocesses:
\begin{eqnarray} \label{700}
\tau_{tr-dwell}=\frac{m}{2\hbar k\kappa^3}\left[(\kappa^2+k^2)\arctan\sqrt{\frac{\kappa}{k}}-\beta\kappa_0^2
\frac{\sqrt{k\kappa}}{k+\kappa}\right];\\
\tau_{ref-dwell}=\frac{2m}{\hbar \kappa^2}\left[\arctan\sqrt{\frac{\kappa}{k}}-\frac{\sqrt{\kappa k}}{k+\kappa}\right].\nonumber
\end{eqnarray}

Thus, if $\tau_{free}= (x_c-a)/v_\kappa$ then $\tau_{tr-dwell}-\tau_{free}$ (which will be denoted as $\tau_{tr-dwell-del}$) is the dwell delay
time for transmission:
\begin{eqnarray} \label{eq.4}
\tau_{tr-dwell-del}=\frac{m(\kappa-k)}{2\hbar k\kappa^3}\left[(\kappa-k)\arctan\sqrt{\frac{\kappa}{k}}+\sqrt{k\kappa}\right].
\end{eqnarray}
This quantity is negative for $V_0>0$, but positive for $V_0<0$. This agrees with the fact that the flow velocity $v_{tr-flow}(x,k)$ in the
transitional region is greater than $v_\kappa$, when $V_0>0$ (see fig.~\ref{fig.1}), but less than $v_\kappa$, when $V_0<0$ (see
fig.~\ref{fig.2}).
\begin{figure}
\begin{center}
\includegraphics[scale=0.2]{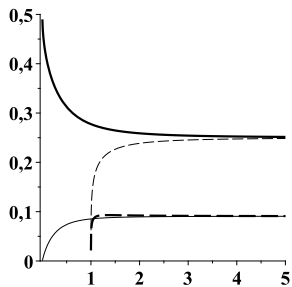}
\caption{$\kappa (x_c-a)/\pi$ for $V_0<0$ (bold solid line) and $V_0>0$ (dashed line) as well as $\kappa l_{depth}/\pi$ for $V_0<0$ (solid line)
and $V_0>0$ (bold dashed line) as functions of $k/\kappa_0$} \label{fig.3}
\end{center}
\end{figure}

As regards reflected particles, by analogy with \cite{Dav} we introduce for them the effective penetration depth. Assuming that reflected
particles move in the region $x>a$ with the velocity $v_k$ and $\tau_{ref-dwell}=2 l_{depth}/v_k$, we obtain
\begin{eqnarray} \label{eq.51}
l_{depth}=\frac{k}{\kappa^2}\left[\arctan\sqrt{\frac{\kappa}{k}}-\frac{\sqrt{\kappa k}}{k+\kappa}\right].
\end{eqnarray}
Thus, the {\it average} turning point for reflected particles lies always to the left of their {\it extreme right} turning point:
$a+l_{depth}(k)\leq x_c(k)$ (see fig~\ref{fig.3}). For $V_0>0$ the penetration depth $l_{depth}(k)$ diverges in the limit $k\to\kappa_0+0$
($\kappa\to 0$) as $\kappa^{-1/2}$. But $l_{depth}=0$ in the limit $k\to0$ for $V_0<0$. In both cases, $\kappa l_{depth}\to(\pi-2)/4$ and $\kappa
(x_c-a)\to\pi/4$ when $k\to \infty$.

So, for a particle scattering at the potential step the flow velocity concept leads in our approach to nonzero transmission and reflection delay
times.

\subsection{Group times for transmission and reflection}

Our next step is to study the temporal aspects of both subprocesses on the basis of the group velocity concept. More precisely, we intend here to
determine asymptotic group times for narrow wave packets peaked at the point $\bar{k}$, where $\bar{k}\neq 0$ when $V_0<0$, and
$\bar{k}\neq\kappa_0$ when $V_0>0$; hereinafter, dealing with such packets, we will omit the top dash in the notation $\bar{k}$.

Before doing so, we want to note that, for studying the time-dependent SWFs (\ref{200}), it is suitable to rewrite the stationary SWFs
$\psi_{tr}(x,k)$ and $\psi_{ref}(x,k)$ in terms of 'usual' solutions of the Schr\"{o}dinger equation, which are continuous everywhere on the $OX$
axis together with their first spatial derivatives. Indeed, let $\Phi_{ref}(x,k)$ be the wave function determined by Exp.~(\ref{208}) in the
region $x<a$ and by Exps.~(\ref{210}) {\it everywhere} in the region $x>a$. Then the SWFs $\psi_{ref}(x,k)$ and $\psi_{tr}(x,k)$ can be written as
$\psi_{ref}(x,k)=\Phi_{ref}(x,k)\theta(x_c(k)-x)$ and $\psi_{tr}(x,k)=\Psi_{tot}(x,k)-\Phi_{ref}(x,k)\theta(x_c(k)-x)$, where $\theta(x)$ is the
Heaviside step function.

Since we are interested in {\it asymptotic} scattering times, let us consider the asymptotically large interval $[0,a+L]$ where $L\gg l_0$ and let
$\psi_{tr}^{inc}(x,t)$ and $\psi_{ref}^{inc}(x,t)$ be wave packets to describe the transmission and reflection subprocesses at the initial stage
of scattering. These are (see (\ref{209}))
\begin{eqnarray*}
\psi_{tr}^{inc}(x,t)=\frac{1}{\sqrt{2\pi}}\int_{-\infty}^{\infty}\mathcal{A}(k)\sqrt{T(k)}\ooa e^{i\left[kx+\beta\left(\lambda(k)-
\pi/2\right)-E(k)t/\hbar\right]}dk\\
\psi_{ref}^{inc}(x,t)=\frac{1}{\sqrt{2\pi}}\int_{-\infty}^{\infty}\mathcal{A}(k)\sqrt{R(k)}\ooa  e^{i\left[kx+\beta
\lambda(k)-E(k)t/\hbar\right]}dk
\end{eqnarray*}
These wave packets differ from the incident one $\Psi_{inc}(x,t)$ (\ref{103}) to describe the whole scattering process. And in the general case
the average departure time $t_{dep}^{tr}$ ($t_{dep}^{ref}$) of transmitted (reflected) particles from the point $x=0$ does not coincide with the
departure time $t_{dep}^{tot}$ of the whole ensemble of particles; in our setting of the problem $t_{dep}^{tot}=0$ (see (\ref{101})). Thus, the
procedure of zero setting, for clocks used in \cite{Dav} for 'measuring' the transmission and reflection times, is justified only in those cases
when $t_{dep}^{tr}$ and $t_{dep}^{ref}$ coincide with $t_{dep}^{tot}$.

As regards the scattered packets $\psi_{tr}^{fin}(x,t)$ and $\psi_{ref}^{fin}(x,t)$, they coincide with the scattered parts of $\Psi_{tot}(x,t)$
(see (\ref{104})): $\psi_{tr}^{fin}(x,t)\equiv\Psi_{tr}(x,t)$ and $\psi_{ref}^{fin}(x,t)\equiv\Psi_{ref}(x,t)$. Thus, in our model and in the CQM,
the average {\it time of arrival} $t_{arr}^{tr}$ of transmitted particles at the point $x=a+L$ {\it is the same}! This is also true for the time
of arrival $t_{arr}^{ref}$ of reflected particles at the point $x=0$.

Note, the definition of the transmitted wave packet $\psi_{tr}^{fin}(x,t)$ is valid at those points $x$ that lie far enough from the turning
points $x_c(k)$ of all waves that constitute this packet. At the first blush, this requirement cannot be fulfilled in the general case because
$x_c(k)\to \infty$ in the limits $k\to 0$ ($V_0<0$) and $k\to \kappa_0+0$ ($V_0>0$). However this is not the case, because $T(k)x_c(k)\to 0$ in
both these limits. The effective length $<x_c>_{tr}-a$ of the transitional region is finite for any transmitted wave packet with the weighted
function $\mathcal{A}(k)$ exponentially decreasing in the limit $|k|\to\infty$. That is, in the general case $\psi_{tr}^{fin}(x,t)$ is defined for
such $x$ that $<x_c>_{tr}+l_0\ll x$.

As in the model of symmetric potential barriers (see~\cite{Ch2,Ch3,Ch4}), calculations carried out for the potential step show that the norms
$\mathcal{T}$ and $\mathcal{R}$ of $\psi_{tr}(x,t)$ and $\psi_{ref}(x,t)$, respectively, take the same constant values at the initial and final
stages of scattering:
\begin{eqnarray*}
<\psi_{tr}^{inc}|\psi_{tr}^{inc}>=<\Psi_{tr}|\Psi_{tr}> =\int_{-\infty}^{\infty}|\mathcal{A}(k)|^2T(k)dk\equiv \mathcal{T}_{as}\\
<\psi_{ref}^{inc}|\psi_{ref}^{inc}>=<\Psi_{ref}|\Psi_{ref}> =\int_{-\infty}^{\infty}|\mathcal{A}(k)|^2R(k)dk\equiv \mathcal{R}_{as}.
\end{eqnarray*}
Moreover, $\mathcal{T}_{as}+\mathcal{R}_{as}=1$. This means that the transmission and reflection subprocesses behave as {\it alternative} ones not
only at the final stage of scattering (when the scattered wave packets do not overlap each other) but also at the initial stage (when the incident
wave packets $\psi_{tr}^{inc}(x,t)$ and $\psi_{ref}^{inc}(x,t)$ interfere with each other). We have to stress that this is true for wave packets
of any width.

However, $\mathcal{T}+\mathcal{R}\neq 1$ at the very stage of scattering, and there are two reasons why the transmission and reflection
subprocesses cannot be considered now as alternative.

One of them has been discussed in \cite{Ch2,Ch3,Ch4} by the example of symmetric potential barriers, where it leads to the nonconservation of the
norm $\mathcal{T}$. Its essence is that the (a)-(d) requirements ensure the balance of the input ($I_{tr}(x,k)|_{x=x_c-0}$) and output
($I_{tr}(x,k)|_{x=x_c+0}$) probability flow densities only in the stationary case, i.e., only for every single wave $\psi_{tr}(x,k)$ in the wave
packet $\psi_{tr}(x,t)$. However, for the packet itself, the interaction between its main 'harmonic' $\psi_{tr}(x,\bar{k})$ and 'subharmonics'
$\psi_{tr}(x,k)$ leads to the imbalance between the time-dependent input and output flow densities at the point $x_c$:
$d\mathcal{T}/dt=I_{tr}(x,t)|_{x=x_c+0}-I_{tr}(x,t)|_{x=x_c-0}\neq 0$.

This effect influences the velocity of the CM of the wave packet $\psi_{tr}(x,t)$. In the case of a narrow packet $\psi_{tr}(x,t)$ peaked at the
point $k$, this influence is maximal when the leading or trailing front of the packet crosses the point $x_c(k)$, i.e., when the role of its
subharmonics is essential. Specifically, when the leading front crosses this point, it serves, depending on the parameters $V_0$ and $k$, as a
source (or sink) of particles. This leads to acceleration (or deceleration) of the CM of the wave packet $\psi_{tr}(x,t)$. When the trailing front
crosses this point, it serves conversely as a sink (or source), leading again to the acceleration (or deceleration) of the CM. Thus, this
acceleration (or deceleration) effect, that has no relation to the average velocity of to-be-transmitted {\it particles}, is accumulated in the
course of the OCS. At the same time the total change of the 'number of particles' involved in the transmission subprocess is strictly zero.

Another reason of violating the equality $\mathcal{T}+\mathcal{R}=1$ is associated with the fact that, unlike symmetric barriers when the function
$x_c(k)$ is constant on the $k$ scale, for the step potential the function $x_c(k)$ is not constant. Now, for the wave packet $\psi_{ref}(x,t)$
with the small width $\Delta k$ ($\Delta k\sim l_0^{-1}$) we have
\begin{eqnarray*}
I_{ref}(x,t)|_{x=x_c(k)-0}\sim \frac{dx_c(k)}{dk}\ooa \Delta k\neq 0,
\end{eqnarray*}
while $I_{ref}(x,t)|_{x=x_c(k)+0}=0$. Due to imbalance between these flows the norm $\mathcal{R}$ now varies at the very stage of scattering (for
symmetric barriers $\mathcal{R}$ is strictly constant (see \cite{Ch2})).

It is evident that, in addition to $\mathcal{R}$, the second effect influences also the norm $\mathcal{T}$. And, unlike the first effect, its
influence on this norm is maximal when the CM of the packet $\psi_{tr}(x,t)$ crosses the point $<x_c>_{tr}$. Of course, the smaller is $\Delta k$,
the smaller is its influence.

Let further $l_0\gg |x^\prime_c(k)/x_c(k)|$ and $x_c(k)-a \ll l_0\ll a+L-x_c(k)$; here the prime denotes the derivative with respect to $k$. For
such narrow packets the second effect is negligible and  $<x_c>_{tr}\approx <x_c>_{ref}\approx x_c(k)$.

In this case the CM's dynamics for both subprocesses is described, at the initial and final stages of scattering, by the expressions
\begin{eqnarray*}
<x>_{tot}^{inc}=\frac{\hbar k}{m}t,\ooo <x>_{tr}^{inc}=<x>_{ref}^{inc}=\frac{\hbar k}{m}t-\beta \lambda',\\
<x>_{tr}^{fin}=\frac{\hbar \kappa}{m}t+a\left(1-\frac{\kappa}{k}\right),\ooa  <x>_{ref}^{fin}=-\frac{\hbar k}{m}t+2a.
\end{eqnarray*}

As it follows from them, the CMs of the narrow wave packets $\psi_{tr}(x,t)$ and $\psi_{ref}(x,t)$ start from the same point
\begin{eqnarray*}
<x>_{tr}^{inc}(0)=<x>_{ref}^{inc}(0)=x_{start}=-\beta \lambda',
\end{eqnarray*}
which differs from the initial position of the CM of the total wave packet: $<x>_{tot}^{inc}(0)=\bar{x}_{tot}(0)=0$ (see (\ref{101})). As a
result, the departure times $t_{dep}^{tr}$, $t_{dep}^{ref}$ and $t_{dep}^{tot}$ of the CMs of the wave packets $\psi_{tr}(x,t)$, $\psi_{ref}(x,t)$
and $\Psi_{tot}(x,t)$, respectively, are determined by the expressions
\begin{eqnarray} \label{400}
t_{dep}^{tr}=t_{dep}^{ref}=t_{dep}=\beta \frac{\lambda'(k)}{v_k}=\frac{m(k-\kappa)}{\hbar(k\kappa)^{3/2}},\ppp t_{dep}^{tot}=0.
\end{eqnarray}
As regards the arrival times $t_{arr}^{tr}$ and $t_{arr}^{ref}$, they are
\begin{eqnarray} \label{eq.9}
t_{arr}^{ref}=\frac{2a}{v_k},\ppp t_{arr}^{tr}=\frac{a}{v_k}+\frac{L}{v_\kappa}.
\end{eqnarray}

Thus, the group transmission and reflection times for the interval $[0,a+L]$ are, respectively, $\Delta t_{tr}=t_{arr}^{tr}-t_{dep}$ and $\Delta
t_{ref}=t_{arr}^{ref}-t_{dep}$. Considering Eqs.~(\ref{400}) and~(\ref{eq.9}), we obtain for them
\begin{eqnarray*}
\Delta t_{tr}=\frac{a}{v_k}+\frac{L}{v_\kappa}-\beta \frac{\lambda'}{v_\kappa} ,\ppp \Delta t_{ref}=\frac{2a}{v_k}-\beta\frac{\lambda'}{v_k}.
\end{eqnarray*}
Excluding expressions related to the 'free' regions $[0,a]$ and $[x_c(k), a+L]$, we obtain the {\it asymptotic} (extrapolated) group times
$\tau_{tr-group}$ and $\tau_{ref-group}$ for transmission and reflection, respectively:
\begin{eqnarray} \label{402}
\tau_{tr-gr}=(x_c-a)/v_\kappa-t_{dep}= \tau_{free}-t_{dep},\ppp  \tau_{ref-gr}=-t_{dep}.
\end{eqnarray}

However, according to the clock model \cite{Dav}, these characteristic times should be defined as follows: $\tau_{tr-gr}^{Dav}=\tau_{free}$ and
$\tau_{ref-gr}^{Dav}=0$. That is, the 'zero setting' procedure in this model is based in fact on the implicit assumption that the departure times
of the CMs of the transmitted and reflected packets coincide with that of the total wave packet, and hence $t_{dep}=0$. This means that the delay
times for both subprocesses are zero in this model. In particular, in line with \cite{Dav}, reflected particles with $E>V_0$ bounce off the step
instantaneously. This takes place even in the limit $E\to V_0+0$ for $V_0>0$.

In our approach the group delay time for the CM of the reflected wave packet is nonzero in the general case. This delay is fully determined by the
group reflection time: $\tau_{tr-gr-del}=\tau_{tr-gr}=-t_{dep}$. For the transmitted CM the delay time is
$\tau_{tr-gr-del}=\tau_{tr-gr}-t_{free}=-t_{dep}$. So that, in the last analysis, for the potential step we obtain
\begin{eqnarray} \label{405}
\tau_{tr-gr-del}=\tau_{ref-gr-del}=-t_{dep}=\frac{m(\kappa-k)}{\hbar(k\kappa)^{3/2}}.
\end{eqnarray}
For $V_0>0$ both group delay times are negative; for $V_0<0$ they are positive.

\subsection{Scattering times for the total reflection} \label{compref}

Let us now consider the case when $V_0>0$ and $E<V_0$, i.e., when transmission is absent: $T(k)\equiv 0$ in the region $(0,\kappa_0]$. Now, in
Exps.~(\ref{eq.1})
\begin{eqnarray*}
A=\frac{2k}{k+i\kappa}e^{(\kappa+ik)a},\ooo B=\frac{k-i\kappa}{k+i\kappa}e^{2ika}
\end{eqnarray*}
where $\kappa=\sqrt{\kappa_0^2-k^2}$. For this one-channel scattering the SWF $\psi_{ref}$ transforms into the total wave function $\Psi_{tot}$,
and $\psi_{tr}\equiv 0$. Thus, $\psi_{ref}(x,k)=e^{ikx}+B e^{-ikx}$ for $x<a$, and $\psi_{ref}(x,k)=A e^{-\kappa x}$ for $x>a$.

Calculations yield that now
\begin{eqnarray} \label{eq.12}
\tau_{ref-dwell}=\frac{\int_a^\infty |\psi_{ref}(x,k)|^2dx}{I_{ref}^{inc}}=\frac{2mk}{\hbar \kappa \kappa_0^2}=\frac{2l_{depth}}{v_k},\ooo
l_{depth}=\frac{k^2}{\kappa\kappa_0^2}.
\end{eqnarray}
As regards the group reflection time, since now $\lambda\equiv 0$ we have $t_{dep}(k)\equiv 0$ and $\tau_{ref-group}=2m/\hbar k\kappa=2/\kappa
v_k$ as in the clock model \cite{Dav}.

On the basis of this expression Davies defines the length scale $d=1/\kappa$ and treats it as "the expectation value for the penetration depth of
the particle into the potential step" \cite{Dav}. But our analysis shows that this interpretation is unjustified.

\section{On the average penetration depth of reflected particles into the potential step}

\begin{figure}
\begin{center}
\includegraphics[scale=0.2]{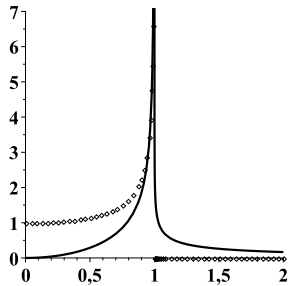}
\caption{The dimensionless penetration depths $\kappa_0l_{depth}$ (solid line) and $\kappa_0 d$ (circles) as functions of $k/\kappa_0$ for
$V_0>0$} \label{fig.4}
\end{center}
\end{figure}

By our approach, only the parameter $l_{depth}$ defined on the basis of the reflection {\it dwell} time can play the role of the average
penetration depth of reflected particles into the potential step. Fig.~\ref{fig.4} shows $l_{depth}$ and $d$ for all values of $k$ when $V_0>0$.
As is seen, both quantities behave equally only in the limit $k\to\kappa_0-0$, when they tend to $\infty$. While, everywhere in the region
$k>\kappa_0$ as well as in the long-wave region $k\to 0$ their properties differ cardinally.

When $k>\kappa_0$ the probability to find (to-be-)reflected particles behind the step is nonzero in our model; this takes place in the
transitional region $[a,x_c(k)]$. But in the model \cite{Dav}, based on the CQM, this probability is strictly zero in the region $x>a$, including
the limiting case $k\to\kappa_0+0$. That is, within the CQM (which does not imply the existence of any transitional region in this scattering
problem), the "penetration depth" $d(k)$ has an infinite discontinuity at the point $k=\kappa_0$ when $V_0>0$. This means that the probability to
find a to-be-reflected particle in the region $x>a$ (at some finite distance from the point $x=a$) changes abruptly when the particle energy is
shifted by an infinitesimal magnitude from $E=V_0-0$ to $E=V_0+0$. Such an abrupt change of the quantum probability contradicts the very spirit of
quantum mechanics. In our model, the penetration depth $l_{depth}(k)$ diverges in both one-side limits, $k\to\kappa_0-0$ and $k\to\kappa_0+0$ (see
fig.~\ref{fig.4}).

Not less strange behavior of $d(k)$ takes place when $k\to 0$. As is seen from fig.~\ref{fig.4}, in this case $d\to 1/\kappa_0$. As is obvious,
this contradicts the fact that the probability to find a reflected particle in the region $x>a$ is strictly zero in this limit! Unlike the 'group'
penetration depth $d$, the 'dwell' penetration depth $l_{depth}$ tends to zero in this case.

\section{Conclusion}

Scattering a quantum particle on the potential step consists of two subprocesses when $E>V_0$. In this case, neither characteristic time nor
expectation value of a physical observable can be calculated within the framework of the CQM, because this model does not 'see' the dynamics of
each subprocess at all stages of scattering. It allows neither direct nor indirect measurement of these quantities. The former is impossible
because of interference between the subprocesses. The latter is impossible since this implies the knowledge of the time evolution of each
subprocess at all stages of scattering.

In this paper we present an alternative model of this process, that uniquely reconstructs the prehistory of the transmitted and reflected
subensembles of particles by their final states. Unlike the CQM this model predicts the existence of a transitional region behind the potential
step, where the probability to find reflected particles is nonzero. This region starts at the point $x=a$ where the potential step is situated and
ends at the point $x_c$ that serves as an extreme turning point for reflected particles. The average velocity of transmitted particles in this
region monotonously changes between the asymptotic values $v_k$ and $v_\kappa$.

We define scattering times for each subprocess -- the dwell time (in the time-independent case) and asymptotic group time (for narrow wave
packets). Among these two concepts only the former allows one to reveal the average times of staying the subensembles of transmitted and reflected
particles in the transitional region.

As it turns out, in the time-dependent case the transmission and reflection subprocesses are not strictly alternative at the very stage of
scattering -- the rule $\mathcal{T}+\mathcal{R}=1$ is violated at this stage. More precisely, this takes place when the leading and trailing
fronts of the narrow wave packet $\psi_{tr}(x,t)$ cross the point $x_c$, i.e., when the role of its 'subharmonics' ($k\neq \bar{k}$) is essential.
In these two cases the velocity of the wave-packet's CM depends not only on the average velocity of particles, but also on the character of
changing the norm $\mathcal{T}$.

Thus, on the one hand, the presented decomposition technique confirms once more the well-known truth -- at the micro-level, quantum probability is
irreducible to classical one; the "either-or" rule $\mathcal{T}+\mathcal{R}=1$, that must be fulfilled for alternative subprocesses from the
viewpoint of classical probability theory, does not hold at the very stage of scattering a quantum particle on the potential step. On the other
hand, it shows that, at the asymptotically large distances from the step, the properties of quantum probability in this scattering problem are
compatible with those of classical one. That is, our quantum-mechanical model respects the correspondence principle.

The following pressing task is to elaborate the Larmor-clock procedure for the potential step, which could serve as a basis for the indirect
measurement of the dwell times of both subprocesses.

\section{Acknowledgments}

This work was supported in part by the Programm of supporting the leading scientific schools of RF (grant No 224.2012.2) for partial support of
this work.


\end{document}